\documentclass[12pt]{article}

\def\he{{herm}}
\def\no{{norm}}
\newcommand{\taub}{\ensuremath{B(\vs|\vt)}}

\newcommand{\fock}[1]{\ensuremath{{\cal F}_{#1}}}
\newcommand{\vs}{\textbf{{\textit{s}}}}
\newcommand{\vsb}{\textbf{\textit{\={s}}}}
\newcommand{\vt}{\textbf{{\textit{t}}}}
\newcommand{\vb}{\textbf{{\textit{v}}}}
\newcommand{\la}{\langle}
\newcommand{\ra}{\rangle}
\newcommand{\ket}[1]{\left|{#1}\rangle\right.}
\newcommand{\bra}[1]{\left.\langle{#1}\right|}
\newcommand{\F}[2]{\p_{\t#1}\p_{\t#2} F}
\newcommand{\BB}[2]{\p_{\s#1}\p_{\s#2}\CB}
\newcommand{\B}[2]{\p_{\s#1}\p_{\s#2} B}
\newcommand{\sumk}{\sum\limits_{k=1}^\infty}
\newcommand{\dfrac}[2]{\displaystyle\frac{#1}{#2}}
\newcommand{\pfrac}[2]{\displaystyle\frac{\p #1}{\p #2}}
\newcommand{\p}{\partial}
\newcommand{\z}[1]{{z_{#1}}}

\newcommand{\zb}[1]{\bar z_{#1}}
\newcommand{\ha}[1] {\ensuremath{a_{#1}}}

\newcommand{\at}[1] {\ensuremath{{\bar a}_{#1}}}
\newcommand{\hb}[1]{\ensuremath{ b_{#1}}}
\newcommand{\hbt}[1]{\ensuremath{{ \bar b}_{#1}}}
\newcommand{\htb}[1]{\ensuremath{{ \bar b}_{#1}}}
\newcommand{\s}[1]{s_{#1}}
\renewcommand{\sb}[1]{\bar s_{#1}}

\renewcommand{\t}[1]{t_{#1}}
\newcommand{\bt}[1]{\bar t_{#1}}
\newcommand{\tb}[1]{\bar t_{#1}}
\newcommand{\pp}[1]{p_{#1}}
\def\CB{{\cal B}}
\def\CC{\ensuremath{{\cal C}}}

\def\CH{{\cal H}}

\def\CZ{{\cal Z}}

\catcode`\@=11
\def\numberbysection{\@addtoreset{equation}{section}
           \def\theequation{\thesection.\arabic{equation}}}

\numberbysection

\begin{document}

\title{\hfill\vbox{\normalsize\hbox{hep-th/0211010}
    \hbox{EFI-02-53}%
    }\\
  \medskip
  \Large Integrability in SFT and new representation of KP
  tau-function} 
\author{ Alexey Boyarsky\footnote{boyarsky@alf.nbi.dk}~\footnote{On leave of
    absence from Bogolyubov ITP, Kiev, Ukraine}\\                        
  {\normalsize \it Niels Bohr Institute} \\
  {\normalsize \it Blegdamsvej 17, DK-2100 Copenhagen, Denmark}\\\\
  Oleg Ruchayskiy\footnote{ruchay@flash.uchicago.edu} \\
    {\normalsize \it Enrico Fermi Institute and Dept. of Physics}\\
    {\normalsize \it University of Chicago} \\
    {\normalsize \it 5640 Ellis Ave. Chicago IL, 60637, U.S.A.}}
\date{October 2002}
\maketitle

\begin{abstract}
  We are investigating the properties of vacuum and boundary states in the CFT
  of free bosons under the conformal transformation. We show that transformed
  vacuum (boundary state) is given in terms of tau-functions of
  dispersionless KP (Toda) hierarchies.  Applications of this approach to
  string field theory is considered. We recognize in Neumann coefficients the
  matrix of second derivatives of tau-function of dispersionless KP and
  identify surface states with the conformally transformed vacuum of free
  field theory.
\end{abstract}
\eject
\section{Introduction}
\label{sec:intro}

Integrability plays an important role in physics. It allows one to go beyond
perturbation theory and to discover properties of one's systems which are
usually inaccessible in any other way. That is why, when one finds a hidden
integrability in the problem it usually means a major breakthrough in it and
promises many new and unexpected results.

Recently the integrability was discovered behind the dynamics of conformal
maps~\cite{minwz,wz,kkmwz}. Namely, it was shown that analytic curves on the
plane can be parametrized by their so called \emph{harmonic moments} (set of
complex variables $\t k,\,k\ge0$). These moments proved to be good coordinates
in the space of such curves.  Evolution of the curve with respect to varying
one of those moments, while keeping the rest fixed, turned out to be described
by dynamical flows of the dispersionless 2D Toda Lattice hierarchy.  As it is
well known, all these flows commute with each other and therefore $t_k$ are
indeed well-defined coordinates in the space of analytic curves.

In particular, it was shown that one can associate the tau-function with the
space of analytic curves. Conformal transformation from a curve to the unit
circle is expressed in terms of second derivatives of this tau-function
calculated at values of its arguments, which coincide with harmonic
moments $\t k$ of the curve at hand. This allowed the authors of~\cite{kkmwz}
to introduce the concept of \emph{tau-function of analytic curves}. In
Section~\ref{sec:digression} we provide an overview of some results from
works~\cite{wz,kkmwz}.

These ideas were applied to several problems in condensed matter: (e.g.
Laplacian growth~\cite{minwz}, Quantum Hall effect~\cite{qhe}) and quantum
field theory (solutions of WDVV equations~\cite{bmrwz}). In each of those
cases integrability helped to obtain new results.

In this paper we are proposing a different realization of the idea of
connection between integrability and conformal maps. We realize a tau-function
as a state, which is the conformal transform of the vacuum of conformal field
theory (CFT) in two dimensions. This is done in the following way.  If one
realizes Fock space of CFT in terms of functions of infinitely many formal
variables $s_k$ and represents creation (annihilation) operators as
multiplication on $\s k$, (differentiation with respect to $\s k$), then the
vacuum is just a constant function.  Conformal transformation of the plane
induces the linear transformation of creation and annihilation operators of
the theory, mixing them in general.  Then the natural question arises: what is
the new vacuum, defined with respect to the new annihilation operators, as a
function of the same variables $s_k$ and conformal transformation (which is
parametrized by harmonic moments $t_k$ of the curve mapped by this
transformation to the unit circle)?

We show in Section~\ref{sec:meaning-function-bs} that this function (called
\taub\ in the paper) is a generating function of the (holomorphic) second
derivatives of the (logarithm of\footnote{In case of dispersionless
  hierarchies logarithm of tau-function is a more natural object than
  tau-function itself. Throughout this paper we will abuse the terminology, by
  calling it just a tau-function.})
tau-function of analytic curve calculated at the point $t_k$. Also, considered
as a function of $s_k$ with $t_k$'s fixed, this function $\taub$ itself is a
(logarithm of) tau-function of dispersionless KP (dKP)
hierarchy~\cite{kodama-carroll,sdiff-kp,takasaki-takebe}. 
As a function of $\s k$ it is quite simple, being quadratic in all variables.
Nevertheless  it is closely related to some other  non-trivial tau-function
in variables $\t k$, namely the tau-function of dKP $F_{herm}$,
given by large $N$ limit of Hermitian one-matrix model. Using the specific
homogeneity condition enjoyed by this function we can identify 
it with our $B(\s k =\t k|t)$.

This gives a new, much simpler and more intuitive, free field realization
of some tau-functions of dKP (compared to the fermion constructions
of~\cite{miwa}) and has immediate applications.

The first application we consider is in the area of String Field Theory
(SFT)~\cite{wittenSFT}.  The language adopted in the present paper is an
adequate one in case of SFT and is indeed widely used there.  One of the basic
objects in the SFT construction are so-called ``Neumann'' coefficients. In
their terms the interaction of the theory (the star-product) can be
defined~\cite{gross,cremmer,samuel,peskinI,peskinII}. They are expressed in
terms of conformal mapping of world-sheets of three interacting strings. Using
Neumann coefficients one can also construct the \emph{surface states} - states
in (boundary) CFT, associated with the given conformal transformation. The
description of such states is an important problem of SFT, as some of them
correspond to D-branes in SFT~\cite{zwiebach}.

We show in Section~\ref{sec:sft} that the ``Neumann coefficients'' associated
with the class of conformal transformations, considered in this paper, are
nothing else but second derivatives of tau-function of dKP hierarchy.  This
implies a number of properties, in particular, the algebraic relation between
the elements of the Neumann matrix. Also we show that our constructed
tau-function $\taub$ \emph{is} a representation of the surface state, and thus
provides new geometrical interpretation for this object.

Finally, we show in Section~\ref{sec:bcft} that the proper generalization of our
construction to the case of CFT of free scalar field with the Dirichlet
boundary conditions allows to find a similar representation (in terms of
conformally transformed boundary state instead of conformally transformed
vacuum) for the tau-function $\CB(\vs,\vsb)$ of dispersionless 2D Toda Lattice
hierarchy (dToda)~\cite{takasaki-takebe}.  This tau-function is related to the
large $N$ limit of the so-called normal matrix model~\cite{zaboron,wz-normal}.

This generalization, being quite straightforward at the first glance, is
non-trivial. Homogeneity conditions for the normal matrix model differ from
those for Hermitian matrix model. Identification of $\CB(t,\bar t)$ with the
``tau-function of analytic curves'' is thus more complicated compared to the
previous case of Hermitian one-matrix model. We will address it
elsewhere~\cite{br2}. Right now we only mention that this construction is
needed for applications to the CFT description of the excitation of Quantum
Hall Droplet and to 2D string theory.  Both these problems are known to be
related to normal matrix model (see e.g.~\cite{qhe} and~\cite{KKK,AKK}).

\section{General Setup}
\label{sec:general-setup}

Consider free chiral scalar field $\phi(w)$ in two dimensions
\begin{equation}
  \label{eq:1}
  \phi(w) = \phi_0 + \hb0 \log w -
  \mathop{\sum\nolimits'}\limits_{k=-\infty}^\infty\left(\frac{\hb k}{ k\,
    w^k} \right)
\end{equation}
This theory is a free CFT. Consider its current $J(w)$
\begin{equation}
  \label{eq:2}
  J(w) = \p\phi(w) = \sum\limits_{k=-\infty}^\infty \frac{\hb k}{w^{k+1}} 
\end{equation}
$\hb k$ obey the usual commutation relations: 
\begin{equation}
[\hb k,\,\hb n] = k\,\delta_{k+n,0}
\label{eq:3}
\end{equation}
$J(w)$ is a primary operator of this CFT with the conformal dimension
$\Delta=(1,0)$, therefore we know that for any conformal transformation $w(z)$:
\begin{equation}
  \label{eq:4}
  \tilde J(z) dz = J(w) dw|_{w=w(z)} 
\end{equation}
We represent $\tilde J(z)$ in the form analogous to~(\ref{eq:2}):
\begin{equation}
  \label{eq:5}
  \tilde J(z)= \sum_{n=-\infty}^\infty\frac{\ha n}{z^{n+1}}
\end{equation}
Commutation relations for the $\ha k$'s are the same as for $\hb k$'s.
We can extract modes of the operator $J(w)$ in the following way\footnote{\label{fn:3}
  Everywhere in this paper we choose orientation of the contour of
  integration such that $\oint_\infty dz\,z^{-1} = 2\pi i$}
:
\begin{equation}
  \label{eq:6}
  b_n = \oint_{\infty} \frac{dw}{2\pi i} w^n J(w)
\end{equation}
By virtue of~(\ref{eq:4}) this can be written as:
\begin{equation}
  \label{eq:7}
  b_n = \oint_{\infty} \frac{dz}{2\pi i} [w(z)]^n \tilde J(z)
\end{equation}
Analogously to the equation~(\ref{eq:6}) we know that
\begin{equation}
  \label{eq:8}
  a_k = \oint_{\infty} \frac{dz}{2\pi i} z^k \tilde J(z)
\end{equation}
Expanding $w(z)$ as a power series in $z$ in eq.~(\ref{eq:7}) we get the
linear transformation from $\hb k$ to $\ha n$.

General form of $w(z)$ which we will consider in this paper is
\begin{equation}
  \label{eq:9}
  w(z) = \frac{z}{r} + \sum_{k\ge0}^\infty\frac{\pp k}{z^k}
\end{equation}
It is \emph{univalent} at infinity i.e. maps region around $z=\infty$ into the
region around $w=\infty$ in the one-to-one manner.  General transformation
will have the following form:
\begin{equation}
  \label{eq:10}
  \hb n = C_{n,n}{\ha n}  + C_{n,n-1}\ha{n-1} + \dots +
  C_{n,0} \ha0 +   \sum_{k>0} C_{n,-k} \ha{-k}, \quad n>0
\end{equation}
\begin{equation}
  \label{eq:11}
  \hb{-n} = \sum_{k=n}^\infty C_{-n,-k} \ha{-k},\quad n>0
\end{equation}
where, $C_{k,n}$ are function of coefficients of conformal
transformation~(\ref{eq:9}). For example, $C_{n,n}=r^{-n}$, $C_{n,n-1} =
n\,\pp0\,{r^{-(n-1)}}$, etc.  General form of $C_{n,k}$ is given by
\begin{equation}
  \label{eq:12}
  C_{n,k} = \oint_\infty \frac{d z}{2\pi i} z^{-k-1} \left(\strut
    w(z)\right)^n\quad  \forall\:n,k
\end{equation}
In particular, $C_{0,k}=\delta_{0,k}$, which means that operator \hb0 does not
transform ($\hb0=\ha0$).  

Given set of operators $\hb k$ with the commutation
relation~(\ref{eq:3}) one can build a Fock space \fock b with the vacuum,
chosen by the condition~(\ref{eq:14}):
\begin{equation}
  \label{eq:14}
  \hb n \ket0_b = 0,\; n\ge0
\end{equation}
and $\hb{-n}$'s acting on it as raising operators and thus building the \fock
b.  Correspondingly we may build a Fock space \fock a, starting from vacuum
$\ket0_a$ and acting on it with operators $\ha{- k}$'s.

Now, the question we want to ask is the following: what kind of transformation
in the Fock space is induced by the transformation~(\ref{eq:10}--\ref{eq:11})?
For example, what would correspond in the space \fock a to the vacuum
state~(\ref{eq:14})?

To answer this question we would like to pick explicit realization of Fock
space \fock a. Let's realize the operators $\ha k$ in the following
way\footnote{
  We do not define the action of operator \ha0 in this Section, because we
  will be working in the subspace of \fock a, where $\ha0\ket\psi =0$ }%
:
\begin{equation}
  \label{eq:15}
  a_n = \pfrac{}{\s n};\quad a_{-n}= n \s n;\quad n>0
\end{equation}
Then Fock space \fock a is the space of functions of infinitely many variables
$\s1,\s2,\dots$ Vacuum $\ket0_a$ is a constant.

Let's find the function, that corresponds in \fock a to the
vacuum~(\ref{eq:14}) understanding operators \hb k\ in terms
of~(\ref{eq:10}--\ref{eq:11}).  This function (which we will look for in the
form $\exp\left(B(s)\right)$) obeys the system of equations:
\begin{equation}
  \label{eq:16}
  \begin{array}{rcl}
    \dfrac1r\dfrac{\p B(\vs)}{\p \s 1}& =& - ( \pp1 \s1+2\pp2\s2+3\pp3\s3+\cdots)\\
    \dfrac1{r^2}\dfrac{\p B(\vs)}{\p \s 2}& =& - \left(\dfrac{2 \pp0}{r}
      \dfrac{\p B(\vs)}{\p \s 1}+  2 \left(\pp0 \pp1 + \frac{\pp2}{r}\right)
      s_1 + 2 \left(\pp1^2 +2\pp0\pp2+ \frac{2\pp3}{r}\right) s_2+\cdots
      \vphantom{\dfrac{\dfrac12}{2}}\right)\\  
    &\dots
  \end{array}
\end{equation}
One can easily integrate equations~(\ref{eq:16}) to get (up to the constant of
integration):
\begin{equation}
  \label{eq:17}
  B(\vs) =  \frac12\sum_{n,k=1}^\infty \s k \s n \pfrac{^2 B(\vs)}{\s k \p \s n}
\end{equation}
where coefficients $\B k n$ \emph{do not depend} on $\s k$ and are expressed
in terms of $r,\pp k$ - coefficients of conformal map~(\ref{eq:9}). For
example one has:
\begin{equation}
  \label{eq:18}
    \frac1r\frac{\p^2 B(\vs)}{\p\s1\p\s k} = - k \pp k
\end{equation}
We put the constant of integration in~(\ref{eq:17}) equal to zero.

\section{Interpretations of Function $B(\vs)$}
\label{sec:meaning-function-bs}

In the Sections that follow we are going to show that the function $B(s)$ can
be identified with the logarithm of the tau-function of dispersionless KP
hierarchy~\cite{sdiff-kp,takasaki-takebe,kodama-carroll}. Note that the
function $B(s)$ depends not only on the formal variables $s_k$ (which
correspond to creation operators $a_{-k}$), but also on the (coefficients of)
conformal transformation $w(z)$. It is necessary for the future interpretation
to describe this dependence explicitly. To do this we will need some basic
information about the ``good coordinates'' in the space of conformal
transformations (or, equivalently, in the space of analytic curves).

\subsection{Digression about the tau-function of analytic curves}
\label{sec:digression}

It was shown in~\cite{minwz,wz,kkmwz} that one can describe the conformal
mappings in the following convenient way. Consider mapping $w(z)$ from the
exterior of the curve \CC\ in the $z$ plane to the exterior of the unit circle
in the $w$ plane, univalent at infinity. General map of this type has the
form~(\ref{eq:9}). Obviously, map $w(z)$ fully describes the curve \CC.
Another way to parameterize the curve \CC\ is by the set of so-called
\emph{harmonic moments} $\t k$ given by
\begin{equation}
  \label{eq:19}
  \t k = \frac1k\oint_\CC \frac{dz}{2\pi i} \bar z\, z^{-k}, \; k>0;\quad \t0 =
  \oint_\CC \frac{dz}{2\pi i} \bar z 
\end{equation}
Set $\t0,\t1,\dots$ (which we will collectively denote by $\vt$) plays the role
of coordinates in the space of the analytic curves.  This means, that the
coefficients of conformal map~(\ref{eq:9}) are actually the functions of them:
$r=r(\vt)$, $\pp k = \pp k(\vt)$. Explicit connection between those two sets
of data for \CC\ is given by the function $F(\vt)$ - so called
\emph{tau-function of analytic curves}~\cite{kkmwz}.

One of the definitions of the function $F(\vt)$ is
\begin{equation}
  \label{eq:20}
  F(\vt)=\frac1{\pi^2}\mathop{\int\int}_{\mathrm{int}\:\CC}
  \log\left|\frac 1z - \frac 1\zeta\right|\, d^2z\,d^2\zeta
\end{equation}
This is a functional that maps space of analytic curves into the complex
numbers. One can consider it as a function of $\t k$.  This function turns out
to be a logarithm of tau-function of dispersionless 2D Toda Lattice hierarchy
(dToda).  This is \emph{the same} tau-function, that can be obtained in the
large $N$ limit of normal matrix model~\cite{zaboron}.  Note, that as a
function of complex variables $\t k$ $F(\vt)$ is usually \emph{not} analytic,
therefore we will think of it as an analytic function of two sets of variables
(plus $\t0$): $F(\vt)=F(\t0;\{\t k\};\{\tb k\})$. Nevertheless we will be
using the notation $F(\vt)$.

Coefficients $\pp k$'s as functions of $\vt$ can be read off the
following useful relation:
\begin{equation}
  \label{eq:21}
  \log\frac{w(z)}{z/r} = -\p_{\t0} D(z) F(\vt)
\end{equation}
where 
\begin{equation}
  \label{eq:22}
  D(z) = \sumk \frac{z^{-k}}k\frac{\p}{\p \t k}
\end{equation}
and $\log r^2 = \p^2_{\t0} F(\vt)$.

As any tau-function, $F(\vt)$ obeys the set of \emph{Hirota identities}.
Hirota identities for dToda can be written in the form~\cite{kkmwz}:
\begin{equation}
  \label{eq:23}
  D(z)D(\zeta)F(\vt) - \frac12\p^2_{\t0^2}F(\vt) = \log\frac{w(z)-w(\zeta)}{z-\zeta}
\end{equation}
which in view of~(\ref{eq:21}) provides the relation between the second
derivatives $\F kn$ ($k,n>0$) and $\F0l$ and one can actually express
\emph{all} second derivatives $\F kn$ in terms of the $\F0l$ (see
e.g.~\cite{kodama-carroll} and discussion in~\cite{bmrwz}). For the discussion 
below we would prefer to rewrite~(\ref{eq:21}),~(\ref{eq:23}) in the form
excluding any reference to the conformal transformation $w(z)$:
\begin{equation}
  \label{eq:24}
  (z-\zeta)e^{D(z)D(\zeta)F} = ze^{-\p_{\t0}D(z)F} - \zeta e^{-\p_{\t0}D(\zeta)F}
\end{equation}

It should be noted that \emph{any} tau-function of dToda hierarchy is also a
tau-function of dKP hierarchy, considered as a function of $\t1,
\t2,\t3,\dots$ only with all other ``times'' (i.e.  $\t0$, $\tb k$) fixed
(c.f.~\cite{takasaki-takebe}). The Hirota identity, which suits better for KP
(and for the purpose of our paper) can be easily derived from~(\ref{eq:23}).
Taking $\zeta\to\infty$ in~(\ref{eq:23}) one gets:
\begin{equation}
  \label{eq:25}
  r w(z)=   z + \pp0 r - \p_{\t1}D(z)F =  z + \pp0 r- \sumk\pfrac{^2
    F(\vt)}{\t1 \p \t k} \frac{z^{-k}}{k}
\end{equation}
and thus
\begin{equation}
  \label{eq:26}
  \exp(D(z)D(\zeta)F) = 1 - \frac{D(z)\p_{\t1}F-D(\zeta)\p_{\t1}F}{z-\zeta}
\end{equation}
Eq.~(\ref{eq:26}) is precisely the Hirota equation for dispersionless KP
hierarchy~\cite{kodama-carroll,takasaki-takebe}.
 
Note, that Hirota equations~(\ref{eq:26}) (correspondingly~~(\ref{eq:24})) are
valid for \emph{any} tau-function of the dKP (correspondingly dToda)
hierarchy, not only for the ``tau-function of analytic curves'' described
above.  So, one can take point of view in a sense opposite to that
of~\cite{wz,kkmwz}. One can \emph{define} some univalent conformal map by
the~(\ref{eq:25}) for KP or~(\ref{eq:21}) for Toda case, for \emph{any}
tau-function of the corresponding hierarchy.\footnote{
  Of course, it is not obvious that for any tau-function $w(z)$, formally
  defined in this way, will have non-zero radius of convergence around
  infinity. But for the wide class of tau-functions it is so. Below when we
  say ``arbitrary'' tau-function we will mean ``any tau-function,
  defining non-trivial conformal map via~(\ref{eq:25}) or~(\ref{eq:21})''. }
In this case one would (in general) lose the interpretation of the times of
dKP (dToda) tau-functions (let's call them $\tilde \t k$) as harmonic moments
of the curve mapped to the unit circle\footnote{
  \label{fn:2}\emph{Different} (slightly more involved) geometrical
  interpretation of the arguments of tau-function $\tilde t_k$ exists for
  \emph{all} tau-functions~\cite{anton}.}
by $w(z)$.  However $\tilde t_k$ would still be coordinates in the space of
conformal maps.  The parameterization of the univalent conformal maps
$w(z|\tilde \t k)$ is then given by the~(\ref{eq:25}) or~(\ref{eq:21}).  If
one uses only Hirota identities (as we will do in the following Sections),
then this parameterization of the conformal maps, given by different
tau-function, can be used as well. However, anything which relies on the
special properties of particular tau-function (which are often derived using
geometrical interpretation, like~(\ref{eq:20})) would not be available any
more.  Example of such an interpretation for the dKP tau-function, related to
the one-matrix model, is discussed in~\cite{mwz}. In their case the contour
gets shrunk to the cut along the real axis and in this limit the tau-function
of analytic curves goes into the partition sum of the Hermitian one-matrix
model.

\subsection{Identification of second derivatives of function $B(\vs)$}
\label{sec:2nd-derivative}

We would like to re-express derivatives $\B kn$ in terms of function $F(\vt)$,
because this would clarify for us the meaning of function $B(s)$. The first
hint that this expression can be very simple is given by comparison of
eq.~(\ref{eq:25}) with eq.~(\ref{eq:18}). Indeed, comparing coefficients in
front of $z^{-k}$ in the right and left hand sides of eq.~(\ref{eq:25}) one
can easily see that
\begin{equation}
  \label{eq:33}
  \pfrac{^2 F(\vt)}{\t1 \p \t k}= -k r \pp k,\;\mbox{ and thus }\; \pfrac{^2
    F(\vt)}{\t1 \p \t k}= \pfrac{^2 B(\vs)}{\s1\p\s k} 
\end{equation}
The natural guess would be that all second derivatives $\B kn$ are equal to
$\F kn$. This is indeed the case, as we show in Appendix~\ref{app:a}.

Thus function $B(\vs)$ becomes a \emph{ generating function} of the matrix of
second derivatives: $\F ij$:
\begin{equation}
  \label{eq:34}
  B(\vs|\vt) = \frac12\sum_{k,n=1}^\infty \s k \s n \frac{\p^2 F(\vt)}{\p t_k
    \p t_n}
\end{equation}
To avoid confusion let's stress once again, that function $B(\vs|\vt)$ is
quadratic function in $\s k$ and its dependence on parameters $\t k$ is
defined entirely by $\p^2_{\t k \t n} F$ and this fact is express by
notations~(\ref{eq:34}).

Let us also mention here, that we did not use yet any properties of the
particular tau-function $F(\vt)$ defined in \cite{wz,kkmwz}. In
Appendix~\ref{app:a} we only used the fact that tau-function satisfies Hirota
equations~(\ref{eq:23}) for the $w(z)$ satisfying~(\ref{eq:21}). The only
place where we implicitly supposed that $F$ is the ``tau-function of analytic
curves'' is where we think about $\t k$ as being harmonic moments of the
curve, mapped by $w(z)$ to the unit circle, i.e. we use the particular
parameterization of conformal maps $w(z|\t k)$. As it was stressed in the
previous Section (see also footnote~\ref{fn:2}, p.~\ref{fn:2}), in principle
we could use equivalent description in terms of different tau-function $\tilde
F(\tilde \t k)$ which would define for us different parameterization of the
conformal maps $w(z|\tilde t_k)$ given again by the same
formulae~(\ref{eq:21}) (or~(\ref{eq:25})).  As a result we would also get an
equivalent to~(\ref{eq:34}) formula for \taub\ (which in this case we call
$B(\vs|\{w\})$ to stress once again its dependence on the conformal mapping
$w(z)$ and not on particular parameterization thereof)
\begin{equation}
  \label{eq:35}
  B(\vs|\{w\})=B(\vs|\tilde t_k) = 
  \frac12\sum_{k,n=1}^\infty \s k \s n \frac{\p^2 \tilde F(\tilde t)}{\p
    \tilde t_k \p \tilde t_n}
\end{equation}
An example of such an equivalent description, which will be of interest for us
here is the one for $\tilde F=F_\he$ and $\tilde t_k=T_k$ . Here $F_{\he}$ is
the tau-function of the dKP hierarchy, equal to the large $N$ limit of the
partition sum of the Hermitian one-matrix model and $T_k$ are new moments
(coupling constants of the matrix model), defined in~\cite{mwz}.  Then we can
rewrite $B(\vs|\{w\})$ in the form
\begin{equation}
 \label{eq:36}
 B(\vs|\{w\})=B(\vs|T_k) = 
 \frac12\sum_{k,n=1}^\infty \s k \s n \frac{\p^2 F_\he(T)}{\p T_k \p T_n}
\end{equation}
In fact, applying the limiting procedure of~\cite{mwz} one can obtain
representation~(\ref{eq:36}) for $B$ directly from~(\ref{eq:34}).

\subsection{Function \taub}
\label{sec:function-b}

The representation~(\ref{eq:34}) means that function \taub\ (as a function of
$\s k$ with all $\t k$ being held constant) satisfies Hirota
equation~(\ref{eq:26}) and thus by itself is the logarithm of tau-function of
dKP hierarchy.

Thus set of equations~(\ref{eq:16}) may be considered as another form of
Hirota identities, because it allows to express all second derivatives of
logarithm of tau-function of dKP in terms of derivatives with respect to
$\t1,\,\t k$.  In the Section~\ref{sec:sft} we will find yet another
interpretation of this function.

One may ask the question then: ``What is the particular condition, which
selects this tau-function among all other tau-functions of dispersionless KP
hierarchy?''.  Let us note, that everything which is said above in this
Section could be valid for any parameterization of the conformal maps given by
some tau-function as it was discussed before. The function $B(\vs)$ as a
function of $\s k$ only will not change if we change this parameterization and
the tau-functions $\tilde F(\tilde t)$.  Note also that equation~(\ref{eq:17})
looks similar to the equation
\begin{equation}
  \label{eq:37}
  \frac12\sum_{k,n=1}^\infty    \t k \t n\pfrac{^2 F_\he^{(0)}}{\t k\p \t n} =
  F_\he^{(0)} 
\end{equation}
where $F_\he^{(0)}$ (called $F_\he$ in the Section~\ref{sec:2nd-derivative}) is
the leading term of the partition sum of Hermitian one-matrix model in the
large $N$ limit (c.f.  Appendix~\ref{sec:homog-herm}). Eq.~(\ref{eq:37}) is
the consequence of the homogeneity condition which is obeyed by particular
tau-function of KP, given by this matrix model (see App.~\ref{sec:homog-herm}
for details).  Thus we are able to identify our tau-function $B(\vs|T)$ with
the one, given by Hermitian one-matrix model in large $N$ limit:
\begin{equation}
  \label{eq:38}
  B(T|T)= \frac12\sum_{k,n=1}^\infty T_k T_n\pfrac{^2 F_\he (T)}{T_k \p
    T_n } = F_\he(T) 
\end{equation}

\section{String Field Theory}
\label{sec:sft}

Let's turn to the application of these ideas now. As mentioned before, the
language of this paper is useful in the completely different field of String
Field Theory (SFT) (c.f.~\cite{peskinI,peskinII}). The subject is huge and
there are many reviews of it (see, e.g.~\cite{SFTreview} and references
therein). Here we briefly remind necessary for us formulae. This is not
intended as an introduction to the subject, but only serves to specify our
notations.

Action of the \emph{Cubic String Field Theory} has the following schematic
form~\cite{wittenSFT}:
\begin{equation}
  \label{eq:39}
  S_{SFT} = \frac12\int \Phi*Q_B\Phi +\frac13\int \Phi*\Phi*\Phi
\end{equation}
We will not discuss the kinetic term here.  Cubic vertex
\begin{equation}
  \label{eq:40}
  V(A,B,C) \equiv \int \Phi_A*\Phi_B*\Phi_C
\end{equation}
can be defined in several ways. The first (so called \emph{operator}) approach
is the following: given three string states $\ket A_1,\ket B_2,\ket C_3$
(corresponding to $\Phi_A,\Phi_B,\Phi_C$), each belonging to its own Hilbert
space $\CH_1,\CH_2, \CH_3$, their interaction vertex $V(A,B,C)$ is given by
\begin{equation}
  \label{eq:41}
  V(A,B,C) = \bra{V_3}\left(\strut\,\ket A \otimes \ket B \otimes \ket C\right)
\end{equation}
where $\bra{V_3}\in\CH_1^*\otimes\CH_2^* \otimes\CH_3^*$ is defined\footnote{
  We write eq.~(\ref{eq:42}) schematically, for one scalar field,
  suppressing ghosts, integration over momenta, etc. (for details see
  e.g.~\cite{gross,cremmer,samuel} or any review in SFT) }
by the expression :
\begin{equation}
  \label{eq:42}
 \bra{V_3} =
 \bra0_1\otimes\bra0_2\otimes\bra0_3\exp\left(-\frac12\sum_{r,s=1}^3
   \sum_{n,m=1}^\infty\alpha^{(r)}_{n} 
   N^{rs}_{n m} \alpha^{(s)}_{m}\right)
\end{equation}
Here $\alpha^{(s)}_{m}$ are modes of the scalar field, indices $r,s=1,2,3$
number Hilbert spaces $\CH_r$ of each of three strings and \emph{Neumann coefficients}
$N^{rs}_{n m}$ are given by
\begin{equation}
  \label{eq:43}
  N_{nm}^{rs} =  \frac1{n\,m}\oint_0 \frac{d z}{2\pi i} z^{-n}\oint_0\frac{d
    \zeta}{2\pi i}\zeta^{-m}\frac{f'_r(z)f'_s(\zeta)}{(f_r(z)-f_s(\zeta))^2}
\end{equation}
where $f_r(z)$ are the conformal transformations of the upper-half plane of
each of the strings to the unit circle.  In terms of $\ket{V_3}$ one can also
define a star-multiplication of any two states:
\begin{equation}
  \label{eq:44}
  \ket{A*B} = \left(\strut\bra{A}\otimes\bra{B}\:\right)\ket{V_3}
\end{equation}

There exists another definition of $V(A,B,C)$ more useful in
applications~\cite{peskinI,peskinII}:
\begin{equation}
  \label{eq:45}
   V(A,B,C) = \la( f_1\circ\Phi_A)(0)\, (f_2\circ\Phi_B)(0)\, (f_3\circ\Phi_C)(0)\ra
\end{equation}
where correlator in the r.h.s. of~(\ref{eq:45}) is computed in any CFT (e.g.
CFT on the unit disk) and the maps $f_i$ ($i=1,2,3$) are the same as in
eq.~(\ref{eq:43}).  Expression in the r.h.s. of~(\ref{eq:45}) has the
following meaning.  One takes (primary) operator $\Phi$ and acts on it with
the conformal transformation $f(z)$ (we will denote the conformal image of
$\Phi$ under the action of $f$ as $(f\circ\Phi)$ or $f[\Phi]$):
\begin{equation}
  \label{eq:46}
  (f\circ\Phi)(z) = f[\Phi(z)] = (f'(z))^d \Phi(f(z))=U_f\Phi(z)U_f^{-1}
\end{equation}
where $d$ is the conformal dimension of the operator. The operator $U_f$ can
be realized in the following way (for $f(z)$ regular at the
origin)\footnote{\label{fn:1}
  We have as usual $L_n\ket0= 0,\;n\ge-1$ and correspondingly $\bra0 L_n =
  0,\; {n\le1}$.}
:
\begin{equation}
  \label{eq:47}
  U_f = \exp(\sum_{n\ge2} v_{n} L_n)
\end{equation}
where $v_n$ are Laurent modes of the function $v(z)=\sum v_n z^{n+1}$,
related to the $f(z)$ in the following way:
\begin{equation}
  \label{eq:48}
  e^{v(z)\p_z} z = f(z)
\end{equation}
One can associate the state $\bra f$, corresponding to any conformal map
$f(z)$, defined as
\begin{equation}
  \label{eq:49}
  \bra{f}\Phi\ra = \bra0 (f\circ \Phi)\ket0 \quad\mbox{for all operators }\Phi
\end{equation}
In particular:
\begin{equation}
  \label{eq:50}
  \bra{f}\dots \alpha_{k_1}\dots \alpha_{k_2}\dots\ket0 = \bra{0}\dots
  f[\alpha_{k_1}]\dots   f[\alpha_{k_2}]\dots\ket{0} 
\end{equation}
where $\alpha_k$ are modes of the expansion of the scalar field $X(z)$ and
vacuum $\ket0$ is defined with respect to them.  The state $\bra f$ can be
represented as
\begin{equation}
  \label{eq:51}
  \bra f = \bra0 U_f
\end{equation}
Indeed, notice  that
\begin{equation}
  U^{-1}_f\ket0 = U_f\ket 0 =\ket0
\label{eq:52}
\end{equation}
for $U_f$ given by~(\ref{eq:47}) and taking into account footnote~\ref{fn:1},
p.~\pageref{fn:1} we get the result~(\ref{eq:50}).  For the purpose of
searching surface states one usually needs to find coefficients $v_n$, thus
one needs to solve the equation~(\ref{eq:48}), which is quite non-trivial
generally.

State~(\ref{eq:51}) has oscillator representation as well. Let's consider free
scalar field $X(z)$. Then
\begin{equation}
  \label{eq:53}
  \bra f = \bra0\exp\left(-\frac12 \sum_{n,m=1}^\infty\alpha_{n} N_{n m}^f
    \alpha_{m}\right)
\end{equation}
Coefficients $N_{n m}^f$ are given by the analog of~(\ref{eq:43}):
\begin{equation}
  \label{eq:54}
  N^f_{n m} =  \frac1{n\,m}\oint_0 \frac{d z}{2\pi i} z^{-n}\oint_0\frac{d
    \zeta}{2\pi i}\zeta^{-m}\frac{f'(z)f'(\zeta)   }{(f(z)-f(\zeta))^2}
\end{equation}
Representation~(\ref{eq:53}) is easy to derive if one considers only two
operators $\alpha_k,\alpha_n$ in eq.~(\ref{eq:50}) and notices that integrand
in the r.h.s. of eq.~(\ref{eq:54}) is just a correlator of $\la f[\p X(z)]f[\p
X(\zeta)]\ra$.

\subsection{Neumann coefficients as second derivatives of tau-function}
\label{sec:neumann-coeff}

Now we would like to repeat this construction for the case at hand.  Namely,
we will consider $w(z)$, given by~(\ref{eq:9}) instead of $f(z)$ and construct
the corresponding surface state $\ket w$.

First, let's introduce the field $\vb(z)$:
\begin{equation}
  \label{eq:55}
  e^{\vb(z)\p_z} z \equiv w(z)
\end{equation}
As $w(z)$ is regular at infinity, the expansion of this field $\vb(z) = \sum
\vb_n z^{n+1}$ has only $n\le1$ modes non-zero. As a result $U_w$,
corresponding to eq.~(\ref{eq:47}) is given by
\begin{equation}
  \label{eq:56}
  U_w = \exp(\sum_{n\le-1} \vb_n L_n)
\end{equation}
and contrary to the property $U_f\ket0=\ket0$ here we have
\begin{equation}
  \label{eq:57}
  \bra0 U_w=\bra0 U_w^{-1}=\bra0
\end{equation}
Let's apply the transformation $U_w$ in the operators in the \fock a. We get
\begin{equation}
  \label{eq:58}
  (w\circ\tilde J)(z) = U_w \tilde J(z) U_w^{-1}
\end{equation}
So, for $ U_w\ha{k}U_w^{-1}\equiv w[\ha k]$ as in previous section, we can
define surface state $\ket w$
\begin{equation}
  \label{eq:59}
      \bra{0}\dots \ha{k_1}\dots \ha{k_2}\dots\ket w = \bra{0}\dots
  w[\ha{k_1}]\dots w[\ha{k_1}] \dots\ket{0}  
\end{equation}
where
\begin{equation}
  \label{eq:60}
  \ket w = U_w^{-1}\ket0=\exp\left(\frac12 \sum_{n,m=1}^\infty\ha{- n} N_{n m}^w
    \ha{-m}\right)\ket0 
\end{equation}
Using the definition (\ref{eq:58}) and operator product expansion for the
current $\tilde J(z)$ one can show (see e.g. \cite{peskinII}) that  Neumann
coefficients  $N_{n m}^w$ here are given by:
\begin{equation}
  \label{eq:61}
  N_{nm}^w = \frac1{n\,m}\oint_\infty \frac{d z}{2\pi i} z^{n}\oint_\infty\frac{d
    \zeta}{2\pi i}\zeta^{m}\frac{w'(z)w'(\zeta)}{(w(z)-w(\zeta))^2}
\end{equation}
One can rewrite this expression in the following form:
\begin{equation}
  \label{eq:62}
  N_{nm}^w = \frac1{n\,m}\oint_\infty \frac{d z}{2\pi i} z^{n}\oint_\infty\frac{d
    \zeta}{2\pi i}\zeta^{m}\p_z\p_\zeta \log\left(
    \frac{w(z)-w(\zeta)}{z-\zeta}\right) 
\end{equation}
Note, that 
\begin{equation}
  \label{eq:63}
  \oint_\infty \frac{d z}{2\pi i} z^{-n}\oint_\infty\frac{d
    \zeta}{2\pi i}\zeta^{-m}\frac1{(z-\zeta)^2} =0 \quad\forall\:n,m>0
\end{equation}
Comparing equation~(\ref{eq:63}) with that of~(\ref{eq:23}) we come to the
conclusion that
\begin{equation}
  \label{eq:64}
  N_{n m } = \frac1{n\,m} \frac{\p^2 F(\vt)}{\p\t n \p\t m }
\end{equation}
This means that Neumann matrix is actually the matrix of second derivatives of 
one function, $F(\vt)$, which is associated with conformal map $w(z)$ in the
way described in the Section~\ref{sec:digression}!

This fact in particular provides a number of relation between the matrix
elements of $N_{nm}$ (see e.g.~\cite{kodama-carroll,bmrwz}) as a consequence
of eq.~(\ref{eq:26}). We show first several of them:
\begin{eqnarray*}
  \label{eq:65}
  N_{22} &=&  N_{13} - \frac 12 N_{11}^2 \\
  N_{23} &=&  N_{14} - N_{11}N_{12}\\
  N_{33} &=& \frac 13 N_{11}^3 -  N_{11} N_{13} - N_{12}^2 +  N_{15} \\
  &\cdots&
\end{eqnarray*}
Of the whole matrix $N_{nm}$ only coefficients $N_{1k}$ are independent! 

We should also mention that this construction is trivially generalized for the
case of conformal transformations, regular at the origin $z=0$, considered in
Section~\ref{sec:sft} (see e.g.~\cite{mwz}). In this case Neumann
coefficients~(\ref{eq:54}) are expressed by the same equation~(\ref{eq:64})
with \emph{the same} tau-function $F$.

\subsection{Surface state as conformally transformed vacuum~$B(\vs)$}
\label{sec:surface-state}

If the representation~(\ref{eq:15}) for operators $\ha{-n}$ is used, surface
state~(\ref{eq:60}) looks identical to the exponential of~(\ref{eq:34}).
Indeed, one can see that ``conformally transformed vacuum'' $\ket0_b$ coincides
with~(\ref{eq:60}), i.e.
\begin{equation}
  \label{eq:66}
  \hb k (a)  \ket{w} = 0
\end{equation}
To see this, note, that expressions $ U_w\ha{k}U_w^{-1}\equiv w[\ha k]$ are
\emph{not} equal to $\hb k$.  Expressions for $\hb k$ are given by the inverse
transformation
\begin{equation}
  \label{eq:67}
  \hb k = U_{w^{-1}}\ha{k}U_{w^{-1}}^{-1}=U_w^{-1}\ha{k}U_w
\end{equation}
Eq.~(\ref{eq:67}) is identical to~(\ref{eq:7}) or equivalently
(\ref{eq:10})--(\ref{eq:11}) i.e. $\hb k$ are just  $w^{-1}[\ha k]$.  
From this and the first formula of~(\ref{eq:34}) one can easily see that
$ b_k \ket w =(U_w^{-1}\ha{k}U_w)\; U_w^{-1}\ket0 = U_w^{-1}~\ha{k}\ket0 =0$. 

We see that the two constructions give equivalent definitions of $\ket w$ and
``surface state'' is nothing else but ``conformally transformed vacuum'' which
was identified in the previous sections with the quadratic tau-function of
integrable hierarchy \taub (or, equivalently, generating function of second
derivatives of the tau-function corresponding to the matrix model).

\section{CFT with the boundary and dToda tau-function}
\label{sec:bcft}

If function \taub\ is actually a logarithm of tau-function of dKP hierarchy,
the question arises - can tau-function for some other hierarchy be obtained in
a similar way. We will demonstrate in this Section that one can obtain the
tau-function of dispersionless 2D Toda Lattice hierarchy in the way similar to
that, taken in
Sections~\ref{sec:general-setup}--\ref{sec:meaning-function-bs}.

Consider the scalar field in the exterior of the unit circle in the plane $w$,
with the Dirichlet boundary conditions on the circle:
\begin{equation}
  \label{eq:68}
    \phi(w,\bar w) = \hb0 \log|w|^2 -
  \mathop{\sum\nolimits'}\limits_{k=-\infty}^\infty\left(\frac{\hb k}{ k\,
    w^k}+\frac{\hbt k}{ k\, {\bar w}^k} \right)
\end{equation}
\begin{equation}
  \label{eq:69}
  \phi(w,\bar w)\Bigr|_{|w|=1} = 0
\end{equation}
Presence of the boundary makes holomorphic and anti-holomorphic modes dependent,
which can be expressed via \emph{boundary state} $\ket\beta$:
\begin{equation}
  \label{eq:70}
  (\hb k - \hbt{-k})\ket\beta=0,\;\forall\:k>0
\end{equation}
Bringing together expressions~(\ref{eq:10}), (\ref{eq:11}) (and their analogs
for the $\htb k$) we get
\begin{equation}
  \label{eq:71}
  \sum_{n=1}^k C_{k,n}\ha n + C_{k,0}\ha0 + \sum_{n=k}^\infty C_{-k,n} \ha{-
    n} -  \sum_{n=k}^\infty {\bar C}_{-k,n} \at{- n} = 0
\end{equation}
where $C_{k,n}$ is given by~(\ref{eq:12}). Similarly
\begin{equation}
  \label{eq:72}
  {\bar C}_{-k,n} = \oint_\infty \frac{d z}{2\pi i} z^{-n-1} \left(\strut
    {\bar w}(z)\right)^{-k},\quad  -\infty< n\le 0,\;k>0
\end{equation}
Again, we realized $\ha k$ as in~(\ref{eq:15}) and introduce new variables
$\sb k$, in terms of which we would realize $\at k$ similarly
to~(\ref{eq:15}), and $\s0$ in terms of which $\ha0$ is realized as
multiplication operator (note, that here, contrary to the case of
Section~\ref{sec:general-setup} boundary state~(\ref{eq:70}) imposes no
restriction on $\hb0$). Then the conformally transformed state in the form
$\exp\left(\CB(\vs,\vsb)\strut\right)$ obeys the equations
\begin{equation}
  \label{eq:73}
  \sum_{n=1}^k C_{k,n}\pfrac{\CB}{\s n} + C_{k,0}\s0 + \sum_{n=k}^\infty
  C_{-k,n} n\s n - \sum_{n=k}^\infty {\bar C}_{-k,n} n \sb n = 0,\quad k>0
\end{equation}
and analogs of~(\ref{eq:73}) where coefficients are conjugated and $\s k$
interchanged with $\sb k$.

Again, one can show (see Appendix~\ref{app:c}) that not only $\p_{\s k}\p_{\s
  n}\CB = \B kn = \F kn$ but also mixed second derivatives $\p^2_{\s k \sb
  n}\CB$ actually coincide with $\p _{\t k}\p_{\tb n}F$.  As a result, we get
the generating function of the matrix of second derivatives~(\ref{eq:74})
$\CB(\s0,\vs,\vsb)$:
\begin{equation}
  \label{eq:74}
  \begin{array}[t]{rcl}
    \CB(\s0,\vs,\vsb) &=& \dfrac{\s0^2}2
  \dfrac{\p^2 F}{\p\t0^2}+\frac12\sum_{k,n=1}^\infty\left(\s k \s n\dfrac{\p^2 F}{\p
      \t k \p \t n}  + \sb k\sb n\frac{\p^2 F}{\p \tb k \p \tb n}\right)+\\[12pt]
  &+&  \s0\sum\limits_{k=1}^\infty \left(\s k \dfrac{\p^2 F}{\p\t0\p\t k} +
    \sb k \frac{\p^2 F}{\p\t0\p\tb k}\right) + \sum\limits_{k,n=1}^\infty \s k
  \sb n\dfrac{\p^2 F}{\p \t k \p \tb n}
\end{array}
\end{equation}
here we've chosen ``constant of integration'' to be $\frac{\s0^2}2
  \frac{\p^2 F}{\p\t0^2}$, so that function $\CB$ would be the generating
  function for \emph{all} second derivatives of dispersionless 2D Toda tau-function.

\subsection{Homogeneity Condition for 2D Toda tau-function}
\label{sec:homog-cond-toda}

Recall that in Section~\ref{sec:function-b} we noticed that function \taub\ 
has the same scaling as partition sum of Hermitian matrix model.  This allowed
us to identify $B(T|T)=F_{herm}(T)$. Let us see if something similar is
possible in the present case.  In the large $N$ limit partition sum of normal
matrix model obeys (see Appendix~\ref{sec:homog-norm-matr}):
\begin{equation}
  \label{eq:75}
  \sumk \left(1-\frac k2\right)\t k \pfrac{F_0}{\t k} + \left(1-\frac
    k2\right)\tb k \pfrac{F_0}{\tb k} + \t0\pfrac{F_0}{\t0} = 2 F_0
\end{equation}
We see that this homogeneity condition has very different than~(\ref{eq:74}).
Thus we can not repeat naively the trick like with the equation (\ref{eq:38}).
This fact is not just a technical detail. It has important mathematical reason
and consequences for physical interpretation.  Detailed discussion of this
issue is beyond the scope of the present paper.  We would like only to note
here that it is important for the application of the method developed here to
the CFT description of the edge excitation of Quantum Hall Effect. We are
going to return to it elsewhere \cite{br2}.
 
\section{Acknowledgments}
\label{sec:Ack}

We would like to thank J.~Harvey, R.~Janik, V.~Kazakov, I.~Kostov, B.~Kulik,
N.~Nekrasov, N.~Obers, K.~Okuyama, A.~Zabrodin, B.~Zwiebach and especially P.~Wiegmann and
J.~Ambjorn for discussions. This work was supported in part by NSF Grant No.
PHY-9901194.  A.B. acknowledges support of Danish Research Council. O.R. would
like to acknowledge the kind hospitality of Niels Bohr Institute where part of
this work was done.

\appendix
\setcounter{equation}{0}
\renewcommand{\theequation}{\Alph{section}.\arabic{equation}}

\section{Computation for holomorphic derivatives}
\label{app:a}

We are going to show that all second derivatives of the function $B(\vs)$ are
expressed through the second derivatives of the function $F(\vt)$.

Taking $n^{th}$ equation in~(\ref{eq:16}) and differentiating it with respect to
$\s k$ one gets:
\begin{equation}
  \label{eq:76}
  \sum_{m=1}^n C_{n,m} \frac{\p^2 B}{\p \s k \p \s m} + k\, C_{n,-k} = 0
\end{equation}
We can rewrite it using definition of~(\ref{eq:12}) as
\begin{equation}
  \label{eq:77}
\sum_{m=1}^{n}\oint_\infty
  \frac{dz}{2\pi i} \frac{ w^n(z) }{z^{m+1}}  \frac{\p^2 B(\vs)}{\p\s
    {m}\p\s  k} =- k \oint_\infty
  \frac{dz}{2\pi i}  w^n(z) z^{k-1},\quad k,n>0
\end{equation}
Obviously, eq.~(\ref{eq:18}) was just a particular case of~(\ref{eq:77}) for
$n=1$. We are going to substitute $\F nm$ into equations~(\ref{eq:77}) and show
that they hold as a consequence of Hirota identity~(\ref{eq:23}).  

First, consider the l.h.s. of~(\ref{eq:77}). Note that as a consequence
of~(\ref{eq:9}) we can substitute $\infty$ instead of $n$ as an upper
summation index in~(\ref{eq:77})
\begin{equation}
  \label{eq:78}
  \sum_{m=1}^n\oint_\infty \frac{dz}{2\pi i} \frac{ w^n(z) }{z^{m}}
  \frac{\p^2 F(\vt)}{\p\t   {m}\p\t  k}= -\oint_\infty \frac{dz}{2\pi i} 
    w^n(z) \p_{\t k} D'(z) F(\vt)
\end{equation}
where we denoted by $D'(z) =\p_z D(z)$, with operator $D(z)$ defined
in~(\ref{eq:22}). Then, taking derivative with respect to $z$ of Hirota
eq.~(\ref{eq:23}) we get:
\begin{equation}
  \label{eq:79}
  \p_{\t k}D'(z)F = \oint_\infty\frac{d\zeta}{2\pi i}
  k\zeta^{k-1}\left(\frac{w'(z)}{w(z) -  w(\zeta)} - \frac1{z-\zeta} \right)
\end{equation}
(recall note~\ref{fn:3}, p.~\pageref{fn:3}).  We assume first that contour of
integration in~(\ref{eq:79}) is chosen so that $|z|>|\zeta|$. Then we can
perform the integration over $\zeta$ in the last term of~(\ref{eq:79}), which
gives zero. Substituting expression~(\ref{eq:79}) back into the r.h.s.
of~(\ref{eq:78}) we get
\begin{equation}
  \label{eq:80}
  -\oint_\infty \frac{dz}{2\pi i} 
  w^n(z) \p_{\t k} D'(z) F(\vt)=
  -\oint_\infty\frac{d\zeta}{2\pi i}
  k\zeta^{k-1}\left(\oint_\infty \frac{dw}{2\pi i} 
    \frac{w^n}{w -  w(\zeta)}\right) 
\end{equation}
We have chosen $|z|>|\zeta|$ which means that $|w(z)|>|w(\zeta)|$ and hence
the contour of integration over $w$ in~(\ref{eq:80}) goes between poles at
$w=\infty$ and $w=w(\zeta)$. The result of integration gives $w^n(\zeta)$
which together with~(\ref{eq:78}) competes the proof\footnote{
  If we have chosen $|z|<|\zeta|$, then the integral in~(\ref{eq:80}) would
  give zero (both poles inside the contour), but the last integral
  in~(\ref{eq:79}) would be equal to $k\zeta^{k-1}$ instead.  So, the final
  result would be of course the same.}.

\section{Computations for mixed derivatives}
\label{app:c}
In case of dispersionless 2D Toda there is another type of Hirota identity,
along with~(\ref{eq:23}) -- the mixed one:
\begin{equation}
  \label{eq:81}
  1-\exp(-D(z)\bar D(\bar \zeta)F) =\frac1{z\bar\zeta}\exp(\p_{\t0}(\p_{\t0} +
    D(z) + \bar D(\bar \zeta))F) 
\end{equation}
The first equation of Hirota~(\ref{eq:81}) is
\begin{equation}
  \label{eq:82}
  \frac{\p^2 F}{\p\t1\p\tb1} = \exp(\p^2_{\t0} F)
\end{equation}
By differentiating~(\ref{eq:82}) with respect to $\t0$ we get the first
equation of dispersionless Toda hierarchy for the function $u$, such that
$\p_{\t0}u=\log r^2 = \p^2_{\t0}F$:
\begin{equation}
  \label{eq:83}
  \pfrac{^2u}{\t1\p\tb1} = \pfrac{}{\t0}\exp\left(\pfrac{u}{\t0} \right)
\end{equation}
Eq.~(\ref{eq:81}) expresses mixed derivatives in terms of derivatives with
respect to $\t0$ and $\t k$ or $\t0$ and $\bt k$.  Now, consider set of
equations~(\ref{eq:73}). First of all, it is obvious that second derivatives
with respect to $\s k, \s n$ obey the same system of equations~(\ref{eq:76})
and thus are derivatives of the function $F$ with respect to the appropriate
harmonic moments $\t k, \t n$. By differentiating eq.~(\ref{eq:73}) with
respect to $\s0$ one can re-write it in the following form:
\begin{equation}
  \label{eq:84}
  \forall\:n>0:\quad\sum_{k=1}^n\frac{\p^2 \CB(\vs,\vsb)}{\p \s0\p \s k} \oint_\infty
  \frac{dz}{2\pi i} 
  w(z)^n z^{-k-1} +  \oint_\infty \frac{dz}{2\pi i} w(z)^n z^{-1}=0
\end{equation}
As in Appendix~\ref{app:a} we would like to substitute $\BB{0}{ k}$ with $\F
0k$ in eq.~(\ref{eq:84}) show that $\F 0k$ obey precisely the same set of
equations, which can be written as (c.f. comment before eq.~(\ref{eq:78})):
\begin{equation}
  \label{eq:85}
  \oint_\infty   \frac{dz}{2\pi i} w(z)^n \p_{\t0}D'(z)F = \oint_\infty
  \frac{dz}{2\pi i} w(z)^n z^{-1}
\end{equation}
where $D'(z)$ was define in Appendix~\ref{app:a}.  As a consequence
of~(\ref{eq:21}) we can write
\begin{equation}
  \label{eq:86}
  -D'(z)\p_{\t0}F = \frac{w'(z)}{w(z)} - \frac1z
\end{equation}
Multiplying l.h.s. of~(\ref{eq:86}) by $w(z)^n$ and integrating around
$z=\infty$ we get:
\begin{equation}
  \label{eq:87}
  \oint_\infty   \frac{dz}{2\pi i} w(z)^n \p_{\t0}D'(z)F =
  -\oint_{\infty}\frac{dw}{2\pi i} w^{n-1} + \oint_\infty
\frac{dz}{2\pi i} w(z)^n z^{-1}
\end{equation}
First terms integrates to zero and we get precisely the r.h.s. of
eq.~(\ref{eq:85})!

Next, we want to show that for mixed derivatives of $\CB$ are equal to those
of $F(\vt)$. To do that, take $n^{th}$ equation~(\ref{eq:73}) and
differentiate it with respect to $\sb k$:
\begin{equation}
  \label{eq:88}
  \sum_{m=1}^n C_{n,m} \frac{\p^2 \CB(\vs,\vsb)}{\p \sb k \p \s m} - k\, \bar C_{-n,-k}
  =0  
\end{equation}
Again, by the same reasons as before, we substitute $\p_{\t k}\p_{\tb n}F$
into~(\ref{eq:88}) and rewrite it as
\begin{equation}
  \label{eq:89}
  \oint_\infty   \frac{dz}{2\pi i} w(z)^n \p_{\tb k}D'(z)F = -k\oint_\infty
  \frac{d z}{2\pi i} z^{k-1} \left(\strut {\bar w}(z)\right)^{-n}
\end{equation}
Rewrite  eq.~(\ref{eq:81}) in the form:
\begin{equation}
  \label{eq:90}
  D(z)\bar D(\zeta)F = -\log\left(1 - \frac1{w(z)\bar w(\zeta)}\right)
\end{equation}
(note, that we should have $|w(z)\bar w(\zeta)|>1$ for the r.h.s. of this
expression to be expansion in $z^{-1}$, $\zeta^{-1}$). Now differentiate it
with respect to $z$ and extract the term, containing $\p_{\tb k}$:
\begin{equation}
  \label{eq:91}
  D'(z) \p_{\tb k} F =  -\oint_\infty   \frac{d\zeta}{2\pi
    i}k\zeta^{k-1}\frac{w'(z)}{w(z)}\,\frac1{w(z)\bar w(\zeta)-1} 
\end{equation}
Now, substituting~(\ref{eq:91}) into the l.h.s. of~(\ref{eq:89}) we get
\begin{equation}
  \label{eq:92}
   \oint_\infty \frac{d\zeta}{2\pi  i}k\zeta^{k-1}\oint_\infty
   \frac{dw}{2\pi i} \frac{w^{n-1}}{w\bar w(\zeta)-1}  =  \oint_\infty
   \frac{d\zeta}{2\pi  i}\frac{k\zeta^{k-1} }{\left(\bar w(\zeta)\right)^n}
\end{equation}
(in the last equation we used the fact that $|w\,\bar w(\zeta)|>1$). This
result coincides with r.h.s. of~(\ref{eq:89}), which proves the statement that
mixed derivatives $\CB(\vs,\vsb)$ are equal to those of $F(\vt)$.

\section{Homogeneity property of Hermitian one-matrix model}
\label{sec:homog-herm}

We show in this Section that partition sum of one matrix model, which is known
to be a tau-function of KP hierarchy~\cite{ivan-cft,KKK} obeys certain
homogeneity condition.  Partition sum can be written as
\begin{equation}
  \label{eq:97}
  \CZ_\he(t) = \int\prod_{k=1}^N d\lambda_k\, \Delta^2(\lambda)\,
  e^{-N V(\lambda)}
\end{equation}
where $V(\lambda)=\sumk \t k \lambda^k$ and $\Delta(\lambda)$ is a
Van-der-Monde. One can see that $\CZ_\he(N;\{\t
k\})=\CZ_\he(N\t1,N\t2,\dots)$. Then if one applies $\sumk\t k \p_{\t k}$ to
$\CZ_\he$ one gets:
\begin{equation}
  \label{eq:98}
  \sumk\t k \pfrac{}{\t k} \CZ_\he(t) = \int\prod_{k=1}^N d\lambda_k\,
  \Delta^2(\lambda)\, e^{-NV(\lambda)}\left(N\sumk \t k
    \lambda^k\right) 
\end{equation}
This result can be also represented as $N\frac{\p}{\p_N} \CZ_\he$. Now, it is
well known (see e.g.~\cite{ginsparg}) that at large $N$ partition sum
of~(\ref{eq:97}) should obey the genus $g$ expansion:
\begin{equation}
  \label{eq:99}
  \log\CZ_\he = \sum_{g=0}^\infty N^{2-2g}F_\he^{(g)}
\end{equation}
So, for $N\to\infty$ property~(\ref{eq:99}) implies: $N\p_N \CZ_\he=2\CZ_\he$.
Thus, for the (logarithm of) partition sum of the Hermitian one-matrix model
one can get in the large $N$ limit
:
\begin{equation}
  \label{eq:100}
  \sumk\t k \pfrac{F_\he^{(0)}}{\t k} = 2 F_\he^{(0)}
\end{equation}
This is precisely the scaling which we need in view of~(\ref{eq:34}).

\section{Homogeneity property of normal matrix model}
\label{sec:homog-norm-matr}

One may wish to repeat the derivation of the Appendix~\ref{sec:homog-herm} for
the case of 2D Toda tau-function $F$, given by the partition sum of normal
matrix model~\cite{zaboron}:
\begin{equation}
  \label{eq:101}
  \CZ_\no = \int \prod_{k=1}^N d^2 \z k\, |\Delta_N(z)|^2\,
  e^{-N V(z,\bar z)}
\end{equation}
where $V(z,\bar z) = -z\bar z + \sumk( \t k \z k + \tb k \zb k)$. The term
$z\bar z$ proves to make a significant difference. Namely, we cannot say that
dependence on $N$ for $\CZ_\no$ enters only in combinations $N\t k$ or $N\tb
k$, there is also an explicit dependence on $N$. To get rid of it, one can
rescale $\z k \rightarrow \z k/\sqrt{N}, \zb k \rightarrow \zb k/\sqrt{N}$ to
\begin{equation}
  \label{eq:102}
  \CZ_\no \to 
  \int \prod_{k=1}^N d^2 \z k\, |\Delta_N(z)|^2\,
  \exp\left(-z\bar z +\sumk\left( \t k N^{1-\frac k2} \z k + \tb kN^{1-\frac k2} \zb k\right) \right)
\end{equation}
Repeating the reasonings similar to those of Appendix~\ref{sec:homog-herm} we
are getting homogeneity condition~(\ref{eq:75}) for $F_0$ --- the leading term
of the large $N$ limit expansion of $\log \CZ_\no$:
\begin{equation}
  \label{eq:27}
  \sumk \left(1-\frac k2\right)\t k \pfrac{F_0}{\t k} + \left(1-\frac
    k2\right)\tb k \pfrac{F_0}{\tb k} + \t0\pfrac{F_0}{\t0} = 2 F_0
\end{equation}
(variable $\t0$ can be introduce in by means of: $\p_{\t0}\CZ_\no = N
\CZ_\no$). We see that the presence of $z\bar z$ terms makes the homogeneity
condition quite different from simply two copies of~(\ref{eq:100}).


\begin{thebibliography}{99}

\bibitem{minwz}M.~Mineev-Weinstein, P.~B.~Wiegmann and
  A.~Zabrodin, ``Integrable structure of interface dynamics,'' Phys.\ Rev.\ 
  Lett.\ {\bf 84}, 5106 (2000) [arXiv:nlin.si/0001007].
  

\bibitem{wz}%
P.~B.~Wiegmann and A.~Zabrodin,
``Conformal maps and dispersionless integrable hierarchies,''
Commun.\ Math.\ Phys.\  {\bf 213}, 523 (2000)
[arXiv:hep-th/9909147].


\bibitem{kkmwz}%
I.~K.~Kostov, I.~Krichever, M.~Mineev-Weinstein, P.~B.~Wiegmann and A.~Zabrodin,
``$\tau$-function for analytic curves,''
arXiv:hep-th/0005259.

\bibitem{miwa}%
  M.~Jimbo and T.~Miwa, ``Solitons And Infinite Dimensional Lie Algebras,''
  Publ.\ Res.\ Inst.\ Math.\ Sci.\ Kyoto {\bf 19}, 943 (1983); \\
  see also: T.~Miwa, M.~Jimbo, E.~Date, \textit{Solitons : differential
    equations, symmetries and infinite dimensional algebras}. Cambridge
  University Press, 2000
  
\bibitem{anton}A.~Zabrodin, ``Dispersionless limit of Hirota equations in some
  problems of complex analysis,'' Theor.\ Math.\ Phys.\ {\bf 129}, 1511 (2001)
  [Teor.\ Mat.\ Fiz.\ {\bf 129}, 239 (2001)] [arXiv:math.cv/0104169].


\bibitem{mwz} A.~Marshakov, P.~Wiegmann and A.~Zabrodin, ``Integrable
  structure of the Dirichlet boundary problem in two dimensions,'' Commun.\ 
  Math.\ Phys.\ {\bf 227}, 131 (2002) [arXiv:hep-th/0109048].

\bibitem{bmrwz} A.~Boyarsky, A.~Marshakov, O.~Ruchayskiy, P.~Wiegmann
  and A.~Zabrodin, ``On Associativity Equations in Dispersionless Integrable
  Hierarchies,'' Phys.\ Lett.\ B {\bf 515}, 483 (2001) [arXiv:hep-th/0105260].


\bibitem{qhe} O.~Agam, E.~Bettelheim, P.~Wiegmann and A.~Zabrodin,
``Viscous fingering and a shape of an electronic droplet in the Quantum Hall regime,''
arXiv:cond-mat/0111333.


\bibitem{br2} A.~Boyarsky, O.~Ruchayskiy ``CFT description of edge excitations
  in Quantum Hall effect from Laughlin functions and integrability'', to be
  published. 

  
\bibitem{sdiff-kp} K.~Takasaki and T.~Takebe, ``SDIFF(2) KP hierarchy,''
  arXiv:hep-th/9112046.
  
\bibitem{ivan-cft}%
  I.~K.~Kostov, ``Conformal field theory techniques
  in random matrix models,'' arXiv:hep-th/9907060.
  
\bibitem{KKK} %
  V.~Kazakov, I.~K.~Kostov and D.~Kutasov, ``A matrix model for
  the two-dimensional black hole,'' Nucl.\ Phys.\ B {\bf 622}, 141 (2002)
  [arXiv:hep-th/0101011].
  
\bibitem{AKK}%
  S.~Y.~Alexandrov, V.~A.~Kazakov and I.~K.~Kostov,
  ``Time-dependent backgrounds of 2D string theory,'' arXiv:hep-th/0205079.


\bibitem{ginsparg}%
  P.~Di Francesco, P.~Ginsparg and J.~Zinn-Justin, ``2-D
  Gravity and random matrices,'' Phys.\ Rept.\ {\bf 254}, 1 (1995)
  [arXiv:hep-th/9306153].
  
\bibitem{kodama-carroll}%
  R.~Carroll and Y.~Kodama, ``Solution of the dispersionless Hirota
  equations,'' J.\ Phys.\ A {\bf 28}, 6373 (1995) [arXiv:hep-th/9506007].

\bibitem{takasaki-takebe}%
K.~Takasaki and T.~Takebe,
``Integrable Hierarchies And Dispersionless Limit,''
Rev.\ Math.\ Phys.\  {\bf 7}, 743 (1995)
[arXiv:hep-th/9405096].

\bibitem{zaboron}%
  L.~L.~Chau and O.~Zaboronsky, ``On the structure of
  correlation functions in the normal matrix model,'' Commun.\ Math.\ Phys.\ 
  {\bf 196}, 203 (1998) [arXiv:hep-th/9711091].

  
\bibitem{wz-normal} %
  P.~Wiegmann and A.~Zabrodin, ``Large scale correlations in normal and
  general nonHermitian matrix ensembles,'' arXiv:hep-th/0210159.

\bibitem{zwiebach}%
  D.~Gaiotto, L.~Rastelli, A.~Sen and B.~Zwiebach, ``Star algebra
  projectors,'' JHEP {\bf 0204}, 060 (2002) [arXiv:hep-th/0202151];\\
  L.~Rastelli and B.~Zwiebach, ``Tachyon potentials, star products and
  universality,'' JHEP {\bf 0109}, 038 (2001) [arXiv:hep-th/0006240];\\
  L.~Rastelli, A.~Sen and B.~Zwiebach, ``String field theory around the
  tachyon vacuum,'' Adv.\ Theor.\ Math.\ Phys.\ {\bf 5}, 353 (2002)
  [arXiv:hep-th/0012251].


\bibitem{SFTreview}%
  K.~Ohmori, ``A review on tachyon condensation in open string field
  theories,'' arXiv:hep-th/0102085.


\bibitem{peskinI}%
  A.~LeClair, M.~E.~Peskin and C.~R.~Preitschopf, ``String Field Theory On The
  Conformal Plane. 1. Kinematical Principles,'' Nucl.\ Phys.\ B {\bf 317}, 411
  (1989).

\bibitem{peskinII}%
  A.~LeClair, M.~E.~Peskin and C.~R.~Preitschopf, ``String Field Theory On The
  Conformal Plane. 2. Generalized Gluing,'' Nucl.\ Phys.\ B {\bf 317}, 464
  (1989).
  
\bibitem{gross}%
  D.~J.~Gross and A.~Jevicki, ``Operator Formulation Of
  Interacting String Field Theory,'' Nucl.\ Phys.\ B {\bf 283}, 1 (1987);
D.~J.~Gross and A.~Jevicki, ``Operator Formulation Of
  Interacting String Field Theory. 2,'' Nucl.\ Phys.\ B {\bf 287}, 225 (1987)

\bibitem{cremmer}E. Cremmer, A. Schwimmer and C. Thorn, ``The Vertex 
  Function in Witten's Formulation of String Field Theory," 
  \textit{Phys. Lett.} \textbf{B179}(1986)57.

\bibitem{samuel}S. Samuel, ``The Physical and Ghost Vertices in 
   Witten's String Field Theory," \textit{Phys. Lett.} 
  \textbf{B181}(1986)255.
  
\bibitem{wittenSFT} E.~Witten, ``Noncommutative Geometry And String
  Field Theory,'' Nucl.\ Phys.\ B {\bf 268}, 253 (1986).

\end{thebibliography}
\end{document}